\documentclass[12pt,a4paper]{article}
\newcommand{\be}{\begin{equation}}
\newcommand{\en}{\end{equation}}
\newcommand{\bea}{\begin{eqnarray}}
\newcommand{\ena}{\end{eqnarray}}

\oddsidemargin .7cm \textwidth 14.5cm \textheight 23cm  \voffset -1cm \topmargin 1cm \fontsize{12pt}{14pt}\selectfont
\usepackage{graphicx}
\begin{document}

\title{\textbf{{A Smarr formula for charged black holes in nonlinear electrodynamics}}}

\author{Leonardo Balart$^{1*}$ and Sharmanthie Fernando$^{2\dagger}$\\   \\ \small $^{1}$ Departamento de Ciencias F\'{\i}sicas, 
Facultad de Ingenier\'{\i}a y Ciencias \\ \small Universidad de La Frontera, Casilla 54-D \\
\small Temuco, Chile\\ \\  \small $^{2}$ Department of Physics, Geology \& Engineering Technology \\ 
\small Northern Kentucky University \\
\small Highland Heights \\
\small Kentucky 41099 \\
\small U.S.A.\\ \\
\emph{\small $^{*}$leonardo.balart@ufrontera.cl} \\
 \emph{\small $^{\dagger}$fernando@nku.edu }}

\date{}

\maketitle

\begin{abstract}
It is well known that the Smarr formula does not hold for black holes in non-linear electrodynamics. The main reason for this is the fact that the trace of the energy momentum tensor for nonlinear electrodynamics does not vanish as it is for Maxwell's electrodynamics. Starting from the Komar integral, we derived a new Smarr-type formula for spherically symmetric static electrically charged black hole solutions in nonlinear electrodynamics. We show that this general formula is in agreement with some that are obtained for black hole solutions with nonlinear electrodynamics.
\end{abstract}

\section{Introduction}	

Smarr formula is a relation connecting the mass of a black hole with other geometrical and dynamical parameters of a black hole, such as spin $J$, charge $Q$, and the electromagnetic potential $\Phi$. For example, the Smarr formula express the mass $M$ of a Kerr-Newman black hole as
\begin{equation}
M = 2 T_H A + 2 \Omega_H J +  \Phi_H q
\,\,\label{smarr-first} \, ,
\end{equation}
where A represents the area of the horizon, $J$ the angular momentum, $T_H$  the Hawking  temperature on the horizon, $\Omega_H$ the angular velocity and $\Phi_H$ the electric potential.
This formula was obtain by applying Euler's theorem for homogeneous functions~\cite{Smarr:1972kt}. In the case of the Reissner-Nordstrom black hole the corresponding Smarr formula simplifies to be 
\begin{equation}
M =  2 T_H A + \Phi_H q
\,\,\label{smarr} \, .
\end{equation}

It is also possible to obtain the Smarr formula using the Komar integral~\cite{Bardeen:1973gs}, which yields
\begin{equation}
M = - \frac{1}{4 \pi} \oint_H dS_{\mu\nu} \nabla^{\mu} \xi^{\nu} + \int_\Sigma dS_\mu (2T^\mu_\nu - T \delta^\mu_\nu )\xi^\nu 
\,\,\label{komar-integral} \, ,
\end{equation}
where $\xi^\mu$ is the time-like Killing vector on the manifold $\mathcal{M}$, $dS_{\mu\nu}$ is the surface element on the horizon $H$, $dS_{\mu}$ denote the element volume on $\Sigma$ which is bounded by the horizon $H$ and the infinity. 

Smarr formula has been obtained for black holes in various theories and dimensions as follows:  Killing symmetries and a Smarr formula for black holes in arbitrary dimensions were studied by Banerjee et.al. \cite{banerjee}. In a similar work, a Smarr formula for rotating black holes in arbitrary dimensions was derived in Refs.~\cite{Myers:1986un, modak}. Smarr relation for $SU(2)$ Einstein-Yang-Mills-dilaton theory was derived by Kleihaus et.al. \cite{Kleihaus:2002tc} and for Einstein-Maxwell-dilaton black holes was derived by Liu et.al. \cite{liu2}. Smarr formulas also have been found for black hole in Einstein-Aether theory \cite{Ding:2015kba}, for Lorentz breaking gravity \cite{conta1}, for Lovelock gravity \cite{Kastor:2010gq, conta2}, and for black holes in three dimensional gravity \cite{btz}. A geometric derivation of the Smarr formula for static AdS black holes was presented in Ref.~\cite{kastor2}. A generalized Smarr formula for charged black holes was obtained from a scaling law in Ref.~\cite{DiazAlonso:2012mb}. Thermodynamics and Smarr formula of black holes in nonlinear electrodynamics including a term P-V have been treated in Refs.~\cite{Azreg-Ainou:2014twa, Azreg-Ainou:2014lua}.

Non-linear electrodynamics has several applications in physics. For example, in quantum-electrodynamics, polarization of the vacuum leads to non-linear effects which do not occur in the tree level Maxwell's electrodynamics. Such interactions are given by the Euler and Heisenberg effective Lagrangian \cite{euler}. Non-linear electrodynamics also has been explored in cosmology. For example, a non-singular FRW cosmology can be obtained by combining a model of non-linear electrodynamics with Einstein's gravity \cite{loren}. In another interesting work, cosmological origins of non-linear electrodynamics has been studied \cite{novello}. Born-Infeld electrodynamics, which was developed initially to cure the divergences of a point charge \cite{born} also has attracted lot of attention due to its relation to string theory \cite{leigh, tsey1, tsey2, berg, gib1}. When non-linear electrodynamics is coupled to Einstein's theory, one can obtain black hole solutions which are regular \cite{fernando1}. The aspects of such black holes will be discussed in detail in section 2.  In black holes arising from non-linear electrodynamics, the light rays do not follow the null geodesics, but, follow a path defined by an effective metric \cite{nora1}. 

Given the importance of non-linear electrodynamics in physics  and black holes, the studies of thermodynamical quantities and expressions such as first law of thermodynamics and Smarr formula take an important place.
The main goal  of the present work is to find a Smarr formula for charged black holes with non-linear electrodynamics sources. Such a formula must account for the fact that the energy-momentum tensor has a non-vanishing trace in non-linear electrodynamics. 

This article is organized as follows. In Section 2, we will given an introduction to black holes in non-linear electrodynamics. In section 3 we will show that charged black hole solutions in non-linear electrodynamics  do not obey the Smarr formula as we know it, but rather an inequality is obtained between the mass $M$ of the black hole and the thermodynamic quantities. We show with three  examples that the inequality has no exclusive direction.  Making use of the Komar integral we obtain an equality, that is to say a Smarr-type formula for such black holes. In Section 4, we apply the Smarr-type formula we obtained to   known cases of non-regular black holes with non-linear electrodynamics, and we consign the respective formulas of the first law that various authors have obtained. In section 5, an interpretation of the Smarr formula obtained will be done. In section 6 we will make comments on the first law of thermodynamics with regard to black holes in non-linear electrodynamics. Finally, in Section 7, conclusions are given.

\section{Black holes in non-linear electrodynamics}

In general in the case of theories of gravity coupled to nonlinear electrodynamics, the action is given by
\begin{equation} 
S = \int d^4x \sqrt{-g}\left[\frac{R}{16 \pi} - L(F)\right]
\,\,\label{action} \, ,
\end{equation}
where g is the determinant of the metric tensor, $R$ is the Ricci scalar and $L(F)$ the Lagrangian density describing non-linear electromagnetic theory dependent on $F = F^{\mu\nu}F_{\mu\nu}/4$. 
There are may black hole solutions in the literature where non-linear electrodynamics is coupled to gravity. One of the well known solutions is the Bardeen black hole ~\cite{Bardeen:1968} which is a regular black hole solution which  does not behave as Reissner-Nordstr\"{o}m black hole asymptotically. Inspired by the Bardeen black hole, Ay\'on-Beato and Garc\'{\i}a in Ref.~\cite{AyonBeato:1998ub} reported another changed black hole which is a regular black hole solution  and asymptotically behaves as Reissner-Nordstr\"{o}m black hole. Other charged black hole solutions are reported in Refs.~\cite{AyonBeato:1999rg, AyonBeato:1999ec, Dymnikova:2004zc, AyonBeato:2004ih, Bronnikov:2000vy, Balart:2014jia, Balart:2014cga, Kruglov:2016ezw, Culetu:2014lca}. These are in essence theories that can be obtained using F-P dual representation~\cite{Salazar:1987ap}, where the electromagnetic Lagrangian is expressed by the function $H(P)$ and the fields $P_{\mu\nu}$ instead of the Lagrangian density $L(F)$ and the electromagnetic field tensor $F_{\mu\nu}$. They are  related by the Legendre transformation $L = P_{\mu\nu}F^{\mu\nu} - H$. However, in what follows we will carry out our calculations in terms of the fields $F$ instead of $P$. Let us note that these Lagrangians $L(F)$ (or $H(P)$) include the parameter $M$ that is associated with the mass of the black hole (see for example  Eq.~(\ref{H-reg}) of the Appendix that is included at the end of this article).

On the other hand, there are also another charged black hole solutions that are obtained by using theories of non-linear electrodynamics, whose Lagrangians, unlike the regular solutions mentioned above, do not depend on the mass parameter $M$ of the black hole. The best known examples of this type are undoubtedly the Born-Infeld Lagrangian~\cite{born, Born:1934} and the Euler-Heisenberg Lagrangian~\cite{Heisenberg:1935qt}. Other examples are given by Refs.~\cite{Soleng:1995kn, 
Hendi:2012zz, Kruglov:2014iwa, Kruglov:2014hpa, Kruglov:2015fcd}.

From the action in Eq.~(\ref{action}),  we obtain the Einstein's field equation
\begin{equation}
R_{\mu\nu} - \frac{1}{2} R g_{\mu\nu} = 8 \pi T_{\mu\nu}
\,\,\label{einstein-eq} \, ,
\end{equation}
where the energy momentum tensor has the form
\begin{equation}
4 \pi T_{\mu\nu} = g_{\mu\nu} L(F) - F_{\mu\alpha} F_{\nu}^{\,\,\alpha} L_F
\,\,\label{tensor-em} \, ,
\end{equation}
here $L_F = dL/dF$.
We consider the following line element for the most general static and spherically symmetric solution
\begin{equation}
ds^2= -f(r) dt^2 + f(r)^{-1}dr^2 + r^2 (d \theta^2+\sin^2\theta d\phi^2)
\,\,\label{elem-gral} \, ,
\end{equation}
where $f(r)$ is an arbitrary function of the coordinate $r$ and the horizons are the roots of the function $f(r)$.
We consider electrically charged solutions, hence, 
\begin{equation}
F_{01} =  -F_{10} = E(r)
\,\,\label{Fmn-em} \, 
\end{equation}
is the only non zero components of the electromagnetic tensor.
Hence
\begin{equation}
F = -\frac{1}{2} E^2
\,\,\label{F-em} \, .
\end{equation}
Hence, the energy-momentum tensor is given by,
\begin{equation}
4 \pi T^0_{\,\,0} = 4 \pi T^1_{\,\,1} = L + E^2 L_F
\,\,\label{T00} \, ,
\end{equation}
\begin{equation}
4\pi T^2_{\,\,2} = 4\pi  T^3_{\,\,3} = L
\,\,\label{T22} \, .
\end{equation}
From the action we can obtain the field equations
\begin{equation}
(F^{\mu\nu} L_F)_{;\nu} = 0    , \,\,\,\,  (^* F^{\mu\nu})_{;\nu} = 0 
\,\,\label{El-f-eq} \, ,
\end{equation}
where $^* F^{\mu\nu}$ is the dual electromagnetic tensor. From the above equations, it follows that,
\begin{equation}
E L_F = \frac{q}{r^2}
\,\,\label{result} \, .
\end{equation}

\section{A Smarr-type formula for charged black holes with nonlinear electrodynamics}

For the Reissner-Nordstrom black hole, which is a solution of Einstein-Maxwell theory, the Smarr formula given in Eq.~(\ref{smarr}) can be derived using the homogeneity of the mass $M$ as a  function of $\sqrt{A}$ and $Q$. In black holes in non-linear electrodynamics, the homogeneity of mass as a function of $\sqrt{A}$ and $Q$ no longer holds and one cannot expect the relation in Eq.~(\ref{smarr}) to hold \cite{Rasheed:1997ns, Breton:2004qa}. In fact we will demonstrate in the following three examples of black holes in non-linear electrodynamics coupled to gravity that indeed the relation does not hold.

If we express the examples in terms of the line element given in Eq.~(\ref{elem-gral}), then $f(r)$  example given in Ref.~cite{AyonBeato:1998ub} is expressed as
\begin{equation}
f(r)= 1-\frac{2}{r}\left(\frac{M r^3}{(r^2+q^2)^{3/2}}- \frac{q^2 r^3}{2(r^2+q^2)^2}\right)  \,\,\label{fn-Ayon-first} \, .
\end{equation}
In this case, it can be shown numerically that
\begin{equation}
M >  2 T_H A + \Phi_H q
\,\,\label{smarr-1} \, ,
\end{equation}
where the electric potential is
\begin{equation}
\Phi_H = \int^\infty_{r_h} E \, dr = \frac{q}{r_h}
\left(\frac{1}{\left(1 + \frac{q^2}{r_h^2}\right)^3} - \frac{3 M r_h}{2 q^2 \left(1 + \frac{q^2}{r_h^2}\right)^{5/2}} \right) + \frac{3 M}{2 q}
\,\,\label{potentail-1} \, .
\end{equation}
The temperature $T_H$ is given by,
\begin{equation}
T_H = \frac{ 1}{ 4 \pi} \left|  \frac{ dg_{tt}} { dr} \right|_{r = r+}
\end{equation}
and the area $A = 4 \pi r_+^2$.

As a second example, if we consider the regular black hole solution reported in Ref.~\cite{Balart:2014jia}, $f(r)$ is given by,
\begin{equation}
f(r)= 1-\frac{2 M}{r}\left( 1 - \frac{q^2}{(q^6 + 8 M^3 r^3)^{1/3}}\right) \,\,\label{alpha3-BH} \, ,
\end{equation}
we obtain the following inequality 
\begin{equation}
M <  2 T_H A + \Phi_H q
\,\,\label{smarr-2} \, .
\end{equation}
Here the electric potential is given by
\begin{equation}
\Phi_H = \int^\infty_{r_h} E \, dr = \frac{q}{r_h}\left(\frac{1 + \frac{3 q^8}{32 M^3 r_h^3}}{\left(1 + \frac{q^8}{8 M^3 r^3_h}\right)^{4/3}}\right)
\,\,\label{potentail-2} \, .
\end{equation}

As third example, lets us consider the Born-Infeld black hole solution given by,
\begin{equation}
f(r) = 1 - \frac{2 M}{r} + \frac{2 b^2}{r}\int_r^\infty \left(\sqrt{r^4 + \frac{q^2}{b^2}} - r^2\right) dr
\,\,\label{B-I} \, ,
\end{equation}
where the electric potential is~\cite{Gunasekaran:2012dq}
 \begin{equation}
\Phi_H = \frac{q}{ r_h} \,
_{2}F_1\left(\frac{1}{4},\frac{1}{2};\frac{5}{4};-\frac{q^2}{b^2r_h^4}\right) 
\,\,\label{potential-B-I} \, ,
\end{equation}

In this case it can be shown numerically that inequality~(\ref{smarr-2}) is also satisfied (see Appendix A).

In contrast to the Maxwell theory, in  nonlinear electrodynamics  the energy-momentum tensor has a non-vanishing trace, which is precisely related to the direction of inequality, as we will observe later.

In particular, when we consider  electrodynamics  of Maxwell theory where the trace of the associated energy-momentum tensor vanishes, we arrive at the expression~(\ref{smarr}). If, on the other hand, we consider a theory with a non-vanishing trace of energy-momentum tensor, as in the case of charged regular
black hole solutions, then we must have an extra term in the expression~(\ref{smarr}), as we shall hereafter show.

Now let's use the Komar integral~\cite{Komar:1958wp} to find a Smarr-type formula for charged regular black holes.  We will  write the Komar integral for the mass evaluated at the infinity boundary as a sum of a integral over a closed surface at the horizon $H$ and a integral on the volume $\Sigma$ which is bounded by the horizon $H$ and the infinity as,
\begin{equation}
M = -\frac{1}{4 \pi} \oint_\infty dS_{\mu\nu} \nabla^{\mu}  \xi^{\nu} = - \frac{1}{4 \pi} \oint_H dS_{\mu\nu} \nabla^{\mu} \xi^{\nu} - \frac{1}{4 \pi} \int_{\Sigma}d S_{\mu} R^{\mu}_{\,\,\nu} \xi^{\nu}
\,\,\label{komar-int} \, ,
\end{equation}
where $dS_\mu$ denote the volume element on $\Sigma$, $dS_{\mu\nu}$ denote the surface element on $H$ and $\xi^{\mu}$ is a time-like Killing vector.

By considering the Einstein's fields equations, and writing the known result for the first integral we obtain
\begin{equation}
M = \frac{\kappa A}{4 \pi} + \int_{\Sigma}d S_{\mu} (2 T^{\mu}_{\,\,\nu} - T \delta^{\mu}_{\nu}) \xi^{\nu}
\,\,\label{komar-int.1} \, .
\end{equation}
where $A$ is the area of the surface at the horizon and $\kappa$ is the surface gravity .

It follows from Eqs.~(\ref{T00}) and~(\ref{T22}) that  
\begin{equation}
4 \pi (2 T^0_0 - \frac{1}{2} T) = 2 L + 2 E^2 L_F - 2 L - E^2 L_F = E^2 L_F = 
\frac{q}{r^2} E
\,\,\label{result-3} \, ,
\end{equation}
and hence
\begin{equation}
M = \frac{\kappa A}{4 \pi} +  \frac{1}{4 \pi} \int_{\Sigma}d S_{\mu} \frac{q}{r^2} E - \frac{1}{2} \int_{\Sigma}d S_{\mu}  T \delta^{\mu}_{\nu} \xi^{\nu}
\,\,\label{komar-int.2} \, .
\end{equation}
Notice that  $E$ is the electric field obtained from the non-linear electrodynamics considered, which asymptotically behaves as  $q/r^2$. Finally
\begin{equation}
M = \frac{\kappa A}{4 \pi} + \Phi_H q - \int_V dV  \omega
\,\,\label{komar-int.3} \, ,
\end{equation}
where $\Phi_H$ is the electric potential. We have followed the notation introduced in Ref.~\cite{Hayward:1997jp} and introduced the parameter   work density $\omega$, given by $\omega  = T/2$.

Returning to the first two solutions mentioned at the beginning of this section,  we can compute $\omega$ and $\int_V dV \omega$. For the solution given by Eq.~(\ref{fn-Ayon-first}), we obtain the following expression for the work density
 \begin{equation}
\omega = \frac{3 M q^2 \left(4 q^4+3 q^2 r^2-r^4\right)+6 q^4 \left(r^2-q^2\right)
   \sqrt{q^2+r^2}}{8 \pi  \left(q^2+r^2\right)^{9/2}}
\,\,\label{w-ayon} \, ,
\end{equation}
and therefore
 \begin{equation}
\int_V dV \omega = \frac{1}{2} \left[M \left(1-\frac{\left(4
   q^2+r_h^2\right)r^3 }{\left(q^2+r_h^2\right)^{5/2}}\right)+\frac{2 q^4
   r_h^3}{\left(q^2+r_h^2\right)^3}\right]
\,\,\label{int-ayon} \, ,
\end{equation}
Here we can numerically show that $\int_V dV \omega < 0$.

For the solution given by Eq.~(\ref{alpha3-BH}), we find that 
 \begin{equation}
\omega = \frac{4 M^4 q^8}{\pi  \left(8 M^3 r^3+q^6\right)^{7/3}}
\,\,\label{w-dec} \, ,
\end{equation}
so that 
 \begin{equation}
\int_V dV \omega = \frac{M q^8}{2 \left(8 M^3 r_h^3+q^6\right)^{4/3}}
\,\,\label{int-dec} \, .
\end{equation}
In the latter case it is straightforward to note that $\int_V dV \omega > 0$.

\section{Application of the result to known cases}

Various expressions of the first law of thermodynamics have been given for charged black holes obtained from a theory of non-linear electrodynamics  where the trace of energy-momentum tensor is non-zero. They are only applicable for some of the solutions that are known. However, they are all consistent with the Smarr formula given by Eq.~(\ref{komar-int.3}). 

\vspace{5mm}

\noindent i) In the Ref.~\cite{Gunasekaran:2012dq}, the authors considered the Born-Infeld black hole solution whose metric function is given by Eq.~(\ref{B-I}). They obtained the first law of thermodynamics in the form $dM = T dS + \Phi_H d q + B_H db$, where 
 \begin{equation}
B_H = \frac{2}{3}b \, r_h^3 \left(1 - \sqrt{1 + \frac{q^2}{b^2 r_h^4}}\,\right) + \frac{q^2}{3 b r_h} \,
_{2}F_1\left(\frac{1}{4},\frac{1}{2};\frac{5}{4};-\frac{q^2}{b^2r_h^4}\right) 
\,\,\label{B-B-I} \, ,
\end{equation}
and $_{2}F_1(a,b;c;z)$ is the Gauss hypergeometric function. Here the respective Smarr formula is $M = 2 T S + \Phi_H  q - B_H b$, which is in agreement with  Eq.~(\ref{komar-int.3}), that is $B_H b = \int_V dV  \omega$, where
 \begin{equation}
\omega = \frac{1}{4 \pi \sqrt{1 + \frac{q^2}{b^2 r^4}}} \left[\frac{q^2}{r^4} -\frac{2 b^2}{r^2}  \left(\sqrt{r^4 + \frac{q^2}{b^2}} - r^2 \right)\right]
\,\,\label{Trace-B-I} \, .
\end{equation}
The authors in Ref.~\cite{Gunasekaran:2012dq} interpreted  the quantity $B_H$ as the vacuum polarization of the Born-Infeld theory.

\vspace{5mm}

\noindent ii) Another expression that has been obtained is presented in Ref.~\cite{Fan:2016hvf}, where the metric function considered is
\begin{equation}
f(r) = 1 - \frac{2 M}{r} - \frac{2 q^3 r^{\mu - 1}}{\alpha(r^\nu + q^\nu)^{\mu/\nu}}
\,\,\label{Fan} \, ,
\end{equation}
where $\mu \geq 3 $ and $\nu > 0$ are dimensionless parameters. 
Here the authors obtain the first law in the form $dM_{ADM} = T dS + \Phi_H d q + \Pi_H d\alpha$, where
\begin{equation}
\Pi_H = \frac{q^3}{4 \alpha^2} \left\{\left[1 + (\mu + 1) \left(\frac{q}{r_h}\right)^\nu\right] \left[1 + \left(\frac{q}{r_h}\right)^\nu\right]^{-\frac{\mu + \nu}{\nu}} - 1\right\}
\,\,\label{Fan-Pi} \, ,
\end{equation}
and 
\begin{equation}
\omega = \frac{\mu \, q^{\nu + 3} \, r^{\mu - 3}}{8 \pi \alpha} \frac{[(1 + \mu) q^\nu + (1 - \nu )r^\nu]}{(r^\nu + q^\nu)^{\frac{\mu}{\nu}+ 2}}
\,\,\label{Fan-omega} \, ,
\end{equation}
and derive the  Smarr formula  to be $M_{ADM} = M + q^3 \alpha^{-1} = 2 T S + \Phi_H q + 2 \Pi_H \alpha$, which is also in agreement with Eq.~(\ref{komar-int.3}).

\vspace{5mm}

\noindent iii) Another case that we can mention corresponds to the gravity model coupled to nonlinear electrodynamics given in Ref.~\cite{Hassaine:2008pw} in four space-time dimensions with
\begin{equation}
f(r) = 1 - \frac{A}{r} + \frac{B}{r^s}
\,\,\label{martinez} \, ,
\end{equation}
where
\begin{equation}
s = \frac{2}{2 p -1}
\,\,\label{martinez-1} \, ,
\end{equation}
and $p$ is a rational number with an odd denominator.
Here the Smarr formula obtained is~\cite{Gonzalez:2009nn} $M = 2 T S + \Phi_H q/p$, which is in agreement with Eq.~(\ref{komar-int.3}) where $\int_V dV  \omega = (1-1/p)\Phi_H q$.

\vspace{5mm}

\noindent iv) Finally, we can mention two other cases for which the respective Smarr formula is obtained in Ref.~\cite{Hendi:2014kha}. One of them corresponds to the solution given in Ref.~\cite{Soleng:1995kn}
\begin{eqnarray}
f(r) &=& 1 - \frac{2 M}{r} - \frac{4 \beta^2 r^2}{3}[\frac{5}{3}\left(\sqrt{1 + \frac{q^2}{\beta^2 r^4}} -1 \right) - \hbox{ln}\left(\frac{1 + \sqrt{1 + \frac{q^2}{\beta^2 r^4}}}{2}\right) \nonumber \\ && - \frac{4}{3}\frac{q^2}{\beta^2 r^4} \, _{2}F_1\left(\frac{1}{4},\frac{1}{2};\frac{5}{4};-\frac{q^2}{\beta^2 r^4}\right) ]
\,\,\label{soleng} \, . 
\end{eqnarray}
Whose work density is 
\begin{equation}
\omega = \frac{\beta ^2}{\pi } \left[\sqrt{\frac{q^2}{\beta ^2 r^4}+1}-2 \, \hbox{ln}\left(\frac{1 + \sqrt{1 + \frac{q^2}{\beta^2 r^4}}}{2}\right)-1\right]
\,\,\label{w-soleng} \, ,
\end{equation}
and therefore
 \begin{eqnarray}
\int_V dV \omega &=&\frac{14}{9} \beta \, r_h^3 \left(1-\sqrt{\frac{q^2}{\beta ^2
   r_h^4}+1}\right)+\frac{4}{3} \beta \, r_h^3 \, \hbox{ln} \left(\frac{1 + \sqrt{1 + \frac{q^2}{\beta^2 r_h^4}}}{2}\right)\nonumber \\ && + \frac{4 q^2}{9 \beta \, r_h} \, _2F_1\left(\frac{1}{4},\frac{1}{2};\frac{5}{4};-\frac{q^2}{\beta
   ^2 r_h^4}\right)
\,\,\label{int-soleng} \, .
\end{eqnarray}

\vspace{5mm}

\noindent v) The other solution considered is inspired by one of the black hole solutions in Ref.~\cite{Hendi:2012zz} with
\begin{equation}
f(r) = 1 - \frac{2 M}{r} - \frac{\beta^2 r^2}{6}\left[1 + \frac{6 q}{\beta r^3}\int dr \left(\sqrt{W\left(\frac{4 q^2}{\beta^2 r^4}\right)} - \frac{1}{\sqrt{W\left(\frac{4 q^2}{\beta^2 r^4}\right)}} \right)          \right]
\,\,\label{hendi} \, 
\end{equation}
where W is Lambert's W function. In this case we obtain the following result
 \begin{equation}
\omega = \frac{\beta}{4 \pi  r^2}  \left[\frac{q \left(W\left(\frac{4 q^2}{\beta ^2
   r^4}\right)-2\right)}{\sqrt{W\left(\frac{4 q^2}{\beta ^2 r^4}\right)}}+\beta 
   r^2\right]
\,\,\label{w-hendi} \, ,
\end{equation}
and
 \begin{equation}
\int_V dV \omega  = \frac{1}{4} \beta  q \left[\frac{r_h \left(W\left(\frac{4 q^2}{\beta ^2
   r_h^4}\right)-1\right)}{\sqrt{W\left(\frac{4 q^2}{\beta ^2 r_h^4}\right)}}+\int
   dr \frac{W\left(\frac{4 q^2}{\beta ^2 r_h^4}\right)-1}{\sqrt{W\left(\frac{4 q^2}{\beta
   ^2 r_h^4}\right)}} \right] + \frac{\beta ^2 r_h^3}{6}
\,\,\label{int-hendi} \, .
\end{equation}

Note that both black hole solutions given above  asymptotically behave as the Reissner-Nordstr\"{o}m solution.
In addition, both solutions obey a Smarr formula given by $M = 2 TS + \Phi_H q - B_H \beta$, which is also in agreement with Eq.~(\ref{komar-int.3}), and where $\beta$ is a nonlinearity parameter and 
\begin{equation}
B_H = \beta^{-1} \int_V dV \omega 
\,\,\label{B-two} \, .
\end{equation}

\section{Interpretation of the additional term in the Smarr formula}

From relations obtained above we can infer two results. The first is that when we calculate $M$ in Eq.~(\ref{komar-int.1}),  there is a part of the trace related to the work due to nonlinear electric potential, so it is the manifestation of a force acting on charged matter. Then the term $\int dV \omega$ also must be a work due to a force acting on charged matter.

It has been proposed in Refs.~\cite{Labun:2008gm,Labun:2008qq} that the effects of trace of the energy-momentum tensor in nonlinear electrodynamics  are analogous to those that produces a cosmological constant. The idea is to separate the energy-momentum tensor into  two parts, that is $T_{\mu\nu} = \overline{T}_{\mu\nu} + g_{\mu\nu} T/4$, where the bar denotes the traceless remainder. Assuming there is a cosmological constant, we can write the Einstein field equations as
\begin{equation}
R_{\mu\nu} - \frac{1}{2}g_{\mu\nu}R = 8 \pi \overline{T}_{\mu\nu} + 8 \pi g_{\mu\nu} (\frac{T}{4} + \frac{\Lambda}{8 \pi})
\,\,\label{einstein.eq.trace}  \, .
\end{equation}  
Hence the second result is that when considering a black hole solution with cosmological constant a Smarr type formula is obtained, where the cosmological constant plays the role of thermodynamical variable, analogous to pressure. The last term in the Eq.~(\ref{komar-int.3}) in a sense  looks like the term $-2 P V$ which appears in the Smarr formula when we consider AdS black holes, where $P = -\Lambda/(8 \pi)$. However, in our case the term $\omega$ depends on the integral. 

On the other hand, we can consider the energy conditions that satisfies the respective energy-momentum tensor as was done in Ref.~\cite{Balart:2016zrd}, where the Bose-Dadhich relation~\cite{Bose:1998uu} was studied for regular black holes~\cite{Balart:2009xr}. Similarly, inequalities given in Eqs.~(\ref{smarr-1}) and~(\ref{smarr-2}) can be explained by the energy conditions the respective energy-momentum tensor of the black hole considered.

In our analysis, let us consider two energy condition: the weak energy condition (WEC) which states that $T^{\mu\nu} t_\mu t_\nu \geq 0$ for all timelike vectors $t_µ$, that is, the local energy density measured by any observer cannot be negative. And the dominant energy condition (DEC) which states that $T^{\mu\nu} t_\mu t_\nu \geq 0$ and $T^{\mu\nu} t_\mu$ must be a non-spacelike vector for all timelike vectors $t_\mu$, or equivalently that $T^{00} \geq |T^{\mu\nu}|$ for each $\mu$, $\nu$, that is, the flow of energy associated with any observer cannot travel faster than light. Note that the DEC includes the WEC.

\begin{figure}
  \centering
		\includegraphics[width=0.7\textwidth]{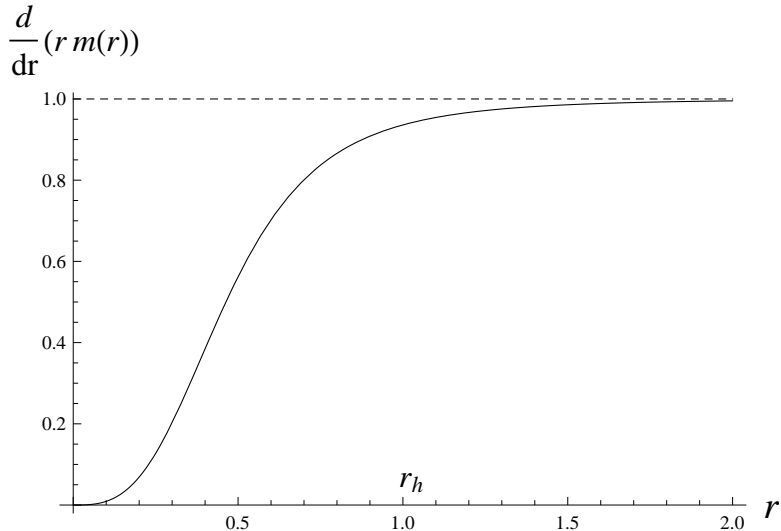}
  \caption{Typical graph of $d(r m(r))/dr$ vs $r$ for regular black hole solution which satisfy the DEC everywhere. Notice that the function $d(r m(r))/dr$ converges from below to M as $r\longrightarrow\infty$. (The numbers on the axes are proportional to M).}
  \label{fig:1}
\end{figure}

If we now consider the line element given by Eq.~(\ref{elem-gral}) with $f(r) = 1 - 2 m(r)/r$, then the components of energy-momentum tensor are~\cite{Dymnikova:2004zc}
\begin{equation}
T^0_{\, \, \, 0} = T^1_{\, \, \, 1} = \frac{2}{8 \pi r^2} \frac{d m(r)}{d r} \,\, , \,\,\,  T^2_{\, \, \, 2} = T^3_{\, \, \, 3}
= \frac{1}{8 \pi r} \frac{d^2m(r)}{dr^2}  \,\,\label{tens-gral} \, ,
\end{equation}
and the energy condition can be expressed in terms of the mass function. Therefore the WEC is equivalent to requiring that the mass function satisfies both
\begin{equation}
 \frac{1}{r^2} \frac{d m(r)}{d r}\geq 0
\,\,\label{1-wec} \, ,
\end{equation}
and
\begin{equation}
\frac{2}{r}\frac{dm(r)}{d r} \geq \frac{d^2m(r)}{dr^2} \,\,\label{2-wec-1-dec} \, .
\end{equation}

Meanwhile the DEC is equivalent to requiring that the mass function satisfies both inequality~(\ref{2-wec-1-dec}) and
\begin{equation}
\frac{2}{r}\frac{dm(r)}{d r} + \frac{d^2m(r)}{dr^2}  \geq 0    \,\,\label{2-dec} \, .
\end{equation}

For its part,
\begin{equation}
\int dV \omega = \int_{r_h}^\infty dr \left(\frac{dm(r)}{d r} + \frac{1}{2}\frac{d^2m(r)}{dr^2}\right) 
= \frac{1}{2}\frac{d(r \, m(r))}{dr}|_{r=\infty} - \frac{1}{2}\frac{d(r \, m(r))}{dr}|_{r=r_h}  \,\,\label{omega-m} \, .
\end{equation}

The results obtained for the examples mentioned, can be expressed in a general way, noting that
\begin{equation}
\frac{d^2 (r \, m(r))}{dr^2} = 2 \frac{dm(r)}{d r} + r \frac{d^2m(r)}{dr^2}  \,\,\label{doubl-deriv} \, ,
\end{equation}
and since we are considering $r \geq 0$, is straightforward to show that if a regular black hole solution satisfies the DEC everywhere, then it satisfies the inequality given in Eq.~(\ref{smarr-2}) or the equality~(\ref{smarr}). This can be  shown in Figure~\ref{fig:1}, where the graph of $d(r m(r))/dr$ vs $r$ allows us to appreciate that when the slope is always positive, i.e. complies with DEC, then $d(r m(r))/dr$ evaluated at infinity is greater than when is evaluated at $r_h$, that is $\int dV \omega \geq 0$.

\begin{figure}
  \centering
		\includegraphics[width=0.7\textwidth]{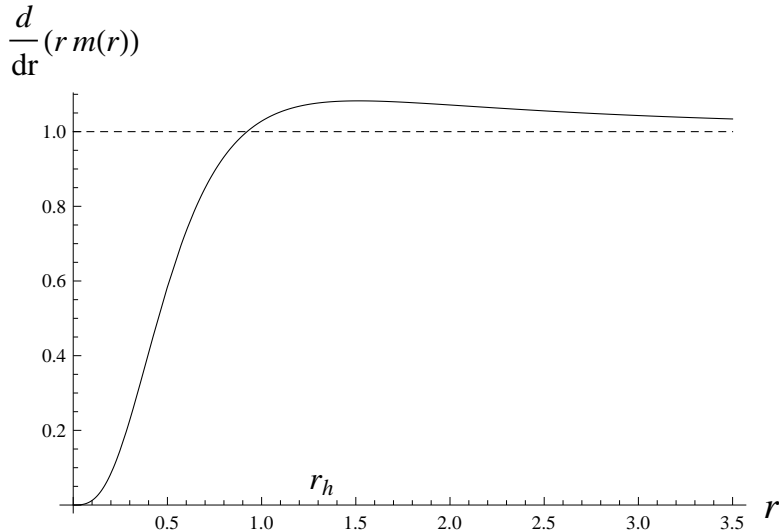}
  \caption{Typical graph of $d(r m(r))/dr$ vs $r$ for regular black hole solution which violate the DEC somewhere but satisfy the WEC everywhere. Notice that the function $d(r m(r))/dr$ converges from above to M as $r\longrightarrow\infty$.}
  \label{fig:2}
\end{figure}

On the other hand, using a similar argument, if the black hole solution complies an inequality as given in Eq.~(\ref{smarr-1}), then the solution violates the DEC in some interval. Particularly for the solution given in Eq.~(\ref{fn-Ayon-first}), which is illustrated in Figure~\ref{fig:2}, we have that $d(r m(r))/dr$ evaluated at $r_h$ is greater than when it  is evaluated at  infinity, which implies that the slope is negative somewhere, that is the DEC is violated in some interval.

In summary, if the energy-momentum tensor of the solution satisfies the DEC everywhere, then it obeys a relation as~(\ref{smarr-2}). And if a solution obeys a relation as~(\ref{smarr-1}), then it violates the DEC somewhere regardless of whether it satisfies the WEC. Note that the implication is to one side only, examples that show that the logical equivalence is not satisfied are given in Ref.~\cite{Balart:2016zrd}.

\section{Comments on  the first law of thermodynamics for black holes in non-linear electrodynamics}

According to the form of its metric function, we can classify the black hole solutions with non-linear electrodynamics in two ways. The first one refers to the solutions whose metric function can be written as follows
\begin{equation}
f(r) = 1 - \frac{2 M}{r} + \frac{q}{r^2} G_1(q, \beta, r)
\,\,\label{metric-G1} \, .
\end{equation}
where  $G_1$ is a function that can be expanded in term of $\beta$, the   nonlinear parameter, which appears in the respective expression of the first law of thermodynamics, that is $dM = T dS + \Phi_H d q + B_H d\beta$.
Black hole solutions~(\ref{B-I}), (\ref{soleng}) and (\ref{hendi}) are examples that illustrate this case.

On the other hand, for  some of the regular black hole solutions obtained with non-linear electrodynamics, the metric function can be written as  
\begin{equation}
f(r) =1 - \frac{2}{r}G_2(M, q, r)
\,\,\label{metric-G2} \, .
\end{equation}
If we want to recover the Reissner-Nordstr\"{o}m solution in the weak field limit, we can not write the function $G_2$ in terms of powers of a certain nonlinearity parameter $\gamma$, because it is unavoidable that it appears in the term proportional to $1/r$ when we make the expansion for $f(r)$, where $f(r)$ becomes
\begin{equation}
f(r)= 1 - \gamma \frac{2 M}{r} + \gamma^2 \frac{q^2}{r^2} + O\left(\frac{1}{r^3}\right)
\,\,\label{metric-serie} \, .
\end{equation}
This is related to the fact that the parameter must be added in the function $H(P)$ and not in the metric function (see Appendix for details). For this reason we can not write the respective first law of thermodynamics in terms of a nonlinearity parameter for metric that have the form given by Eq.~(\ref{metric-G2}).

Black hole solutions~(\ref{Fan}), (\ref{martinez}) are special cases. In the first one the parameter $\alpha$ is a factor that multiplies a function similar to $G_1$, but which does not depend on a non-linear parameter. And in the second the function is also similar to $G_1$, although it has only one term, which is proportional to $1/r^s$.

\section{Conclusions}

In this paper we have derived a new Smarr-type formula for black holes in non-linear electrodynamics. We started the derivation with the Komar integral and obtained a formula which looks similar to the Smarr formula for black holes in Maxwell's electrodynamics plus and additional term. The additional term is related to the non-zero trace of the energy-momentum tensor of the theory of non-linear electrodynamics considered. We have applied the derived formula for several well known black hole solutions.

An interpretation of the additional term in the derived formula is given. A detailed discussion is also included as to how the Dominant energy condition and the Weak energy condition plays a role in the inequalities arrived for some of the black holes for the Smarr formula.

\appendix

\section{Proof of inequality~(\ref{smarr-2}) for the Born-Infel black hole solution}

\begin{figure}[h]
  \centering
		\includegraphics[width=0.7\textwidth]{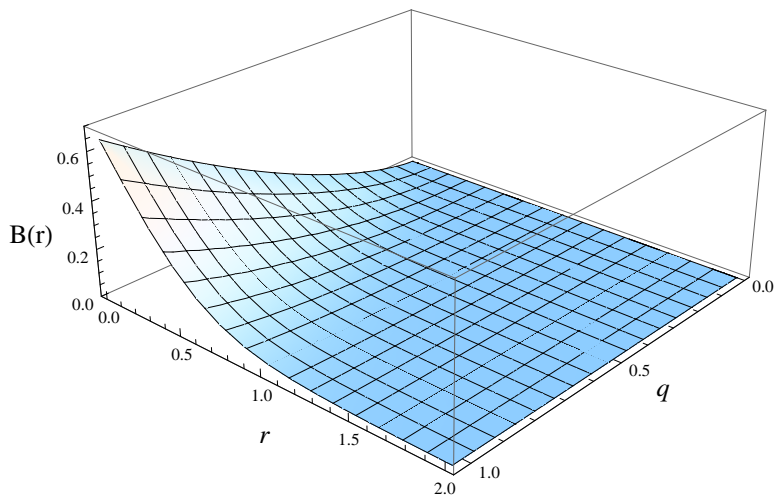}
  \caption{Graph of $B(r)$ vs $r$ for $0 \leq q \leq 1.03 M$  and b = 1.}
  \label{fig:3}
\end{figure}

We can show the inequality~(\ref{smarr-2}) for the Born-Infeld black hole solution in a numerical way considering $B(r)$ defined by Eq.~(\ref{B-B-I}). Figure 3 shows clearly that $B(r) > 0$, even when $r \not= r_h$, and since $M = 2 T S + \Phi_H  q - B_H b$, then the inequality is fulfilled. We obtain the same conclusion for other values of $b$. Note that $q = 1.03 M$ corresponds to the case of extremal black hole for $b = 1$.

\section{Nonlinearity parameter and regular charged black holes}

Let us consider $H(P)$ for the regular black hole solution given by Eq.~(\ref{alpha3-BH}). Now we introduce arbitrary constants $a_1$, $a_2$ and $a_3$, some of which could serve as a nonlinearity parameter
\begin{equation} 
H(P) = \frac{a_1 P}{(a_2 + a_3 \, \Omega \, (-P)^{3/4})^{4/3}}
\,\,\label{H-reg} \,   ,
\end{equation}
where 
\begin{equation} 
P = -\frac{q^2}{2 r^4}
\,\,\label{P-H-reg} \, 
\end{equation}
and  
\begin{equation} 
\Omega  = \frac{2^{3/4} q^{9/2}}{8 M^3}
\,\,\label{Omega} \,   .
\end{equation}
Using the relation 
\begin{equation} 
\frac{d m(r)}{dr} = - r^2 H(P)
\,\,\label{relat-H-m} \, ,
\end{equation}
of the F-P dual formalism, we obtain the respective mass function 
\begin{equation}
m(r)= M \frac{a_1}{a_2} \left(\frac{1}{a_3^{1/3}} - \frac{q^2}{(a_3 \, q^6 + 8 \, a_2\,  M^3 r^3)^{1/3}}\right) \,\,\label{mass-with-ctes} \, ,
\end{equation}

If we take the weak field limit, the following expansion is obtained for the metric  function 
\begin{equation}
f(r)= 1 -  \frac{a_1}{a_2 \, a_3^{1/3}}\frac{2 M}{ r} + \frac{a_1}{a_2^{4/3}} \frac{q^2}{r^2} -  \frac{a_3}{a_2} \frac{q^8}{24 M^3 r^5} + O(r^{-8})\,\,\label{serie-f-ctes} \, .
\end{equation}
In all the terms of higher orders there is a power of $a_3/a_2$. From here, it is now straightforward to see that in order to obtain the Reissner-Nordstr\"{o}m solution in the weak field limit it is imperative that $a_1 = a_2^{4/3}$, and therefore $a_2 = a_3$. Accordingly, if we want to recover the Reissner-Nordstr\"{o}m solution in the weak field limit, it is not possible to include a nonlinearity parameter in the function $H(P)$.

\section*{Acknowledgments}

L. B. is supported by the ``Direcci\'on de Investigaci\'on de la Universidad de La Frontera'' (DIUFRO)  through the project: DI16-0075.

\section*{Note}
After uploading our paper to arXiv, we were informed that there is some overlap between our paper and the paper in Ref. \cite{Gulin:2017ycu}.

\end{document}